# New Metallicity Calibration Down to [Fe/H] = –2.75 dex

**KARAALİ, S., BİLİR, S., KARATAŞ, Y., and AK, S. G.**

*Istanbul University Science Faculty Department of Astronomy and Space Sciences, 34452 Istanbul – TURKEY*

E-mail: karsa@istanbul.edu.tr

**Abstract**

88 dwarfs, covering the colour – index interval $0.37 \leq (B-V)_o \leq 1.07$ mag, with metallicities $-2.70 \leq [Fe/H] \leq +0.26$ dex have been taken from three different sources for new metallicity calibration. The catalogue of Cayrel de Stroble et al. (2001) which includes 65% of the stars in our sample supplies detailed information on abundances for stars with determination based on high–resolution spectroscopy. 77 stars which supplies at least one of the following conditions have been used as "*corner stones*" for constructing the new calibration: (*i*) the parallax is larger than 10 *mas* (distance relative to the Sun less than 100 pc) and the galactic latitude is absolutely higher than $30^o$, (*ii*) the parallax is rather large, if the galactic latitude is absolutely low and vice versa. Contrary to the previous investigations, a third-degree polynomial fitted for the new calibration: $[Fe/H] = 0.10 - 2.76\delta - 24.04\delta^2 + 30.00\delta^3$. The coefficients were evaluated by the least-square method, without regarding the metallicity of Hyades. However, the constant term is in the range of metallicity determined for this cluster i.e.: $0.08 \leq [Fe/H] \leq 0.11$ dex. The mean deviation and the mean error in our work are equal to those of Carney (1979), for $[Fe/H] \geq -1.75$ dex where Carney's calibration is valid.

**Introduction**

Metallicity plays an important role in the Galactic structure. Although mean metal−abundances were attributed to three main Galactic components, i.e.: Population I (Thin Disk), Intermediate Population II (Thick Disk), and Extreme Population II (Halo) (cf. Norris 1996), recent works show that the metallicity distributions for these populations may well be multimodal (Norris 1996, Carney 2000, Karaali et al. 2000). More important is the metallicity gradient cited either for populations individually or for a region of the Galaxy. Examples can be found in Reid & Majewski (1993) and Chiba & Yoshii (1998). The importance is related to the formation of the Galaxy as explained here by. The existence of a metallicity gradient for any component of the Galaxy means that it formed by dissipative collapse. The proponents of this suggestion are Eggen, Lynden−Bell, & Sandage (1962, ELS). Discussion of the current status of this model is provided by Gilmore, Wyse, & Kuijken (1989). Later analyses followed (e.g. Yoshii & Saio 1979; Norris, Bessel, & Pickles 1985; Norris 1986; Sandage & Fouts 1987; Carney, Latham, & Laird 1990; Norris & Ryan 1991; and Beers & Sommer−Larsen 1995). From their studies, an alternative picture emerged, suggesting that the collapse of the Galaxy occured slowly. This picture was postulated largely on a supposed wide age range in the globular cluster system (Searle & Zinn (1978, SZ), Schuster & Nissen (1989). SZ especially argued that the Galactic halo was not formed as a result of collapse but from the merger or accretion of numerous fragments such as dwarf−type galaxies. Such a scenario indicates no metallicity gradient or younger and even more metal−rich objects at the outermost part of the Galaxy. The globular cluster age range supposition has been disproved by recent analyses (Rosenberg et al. 1999) while the number of young field halo stars has been shown to be extremely small, inconsistent with this model by Unavane, Wyse & Gilmore (1996), Preston & Sneden (2000), and Gilmore (2000).

A clear metallicity gradient is highly dependent on the precise metallicity determination. The ultraviolet excess provides metallicities for large field surveys in many photometries such as uvby-β (Strömgren 1966), VBLUW (Walraven & Walraven 1960, Trefzger et al. 1995), RGU (Buser & Fenkart 1990) and UBV (Carney 1979). There are many calibrations between the normalized ultraviolet excess $(\delta_{U-B})_{0.6}$ and the metal abundance [Fe/H] for the last system which deviate from each other considerably. Fig.15 of Buser & Kurucz (1992) compares these calibrations based on empirical data (Cameron 1985, Carney 1979) or theoretical models (Buser & Kurucz 1978, 1985, and Vandenberg & Bell 1985). The reason of these differences originates from the UBV data as well as from the atmospheric parameters. Cayrel de Strobel et al. (2001) state the following discrepancies even for high quality observations and careful analysis:

1) The [Fe/H] determinations are usually solar scaled however it logarithmically changes from author to author by 0.20.

2) The difference in temperature proposed by different authors for a star may be as large as 400 °K which results ≅0.80 dex in metallicity as in the case of the metal–poor halo sub–giant HD 140283.

3) The great metal deficiency of the atmosphere for a very evolved Population II star results its misclassification in spectral type. The MK spectrum of such stars mimics the MK spectrum of a hotter unevolved star.

4) It was shown by Hipparcos data that spectroscopic gravities, based on ionisation equilibrium are in error for very metal–poor stars.

We aimed to derive a new metallicity calibration for stars with large scale in [Fe/H] making use of the updated UBV and [Fe/H] data and keeping in mind the reservations mentioned above. Thus we had to investigate the metallicity distribution of metal-poor stars at different distances from the galactic plane and contribute to the implications for the Galactic formation and evolution. The first application (Karaali et al. 2002) based on the CCD data, for stars in an intermediate latitude field is promising. The data are presented in Section 2. The new metallicity calibration is given in Section 3 and finally a short discussion is presented in Section 4.

## 2. The Data

The data given in Table 1 were taken from three different sources. (1) 57 of them with log g ≥ 4.5 are from Cayrel de Strobel et al. (2001). This catalogue supplies detailed information on abundances for stars with determination based on high–resolution spectroscopy. Also it contains the errors in atmospheric parameters, i.e.: $T_{eff}$, log g, and [Fe/H] when available. However, we did not include such stars in the statistics of our work. Additionally the spectral types of the stars are available in the catalogue. (2) 11 high mass stars were taken from a different catalogue of the same authors (Cayrel de Strobel et al. 1997). This catalogue has the advantage of including metal–poor stars down to [Fe/H] = –2.70 dex with smaller surface gravity, i.e.: log g ≥ 4.0.

We consulted the specialised catalogues which are included in the General Catalogue of Photometric Data[1] (Mermilliod et al. 1997) for the UBV–magnitudes and colours in these catalogues. The data in columns 11 and 12, i.e.: the parallax and the galactic latitude in Table 1, are provided from the SIMBAD database. (3) We selected 20 dwarfs from the catalogue of Carney (1979). Although Table 5 of Carney includes a large sample of dwarfs we had to eliminate 8 of them which are common in the catalogues of Cayrel de Strobel et al. (1997, 2001) and 12 of them which turned out to be variable stars according to the SIMBAD database. $T_{eff}$ and log g parameters not given in the table of Carney are from the authors cited in the "*remarks*" column.

The selection of total 88 stars from the catalogues mentioned above carried out as follows; most of the 57 and 20 stars taken from Cayrel de Strobel et al. (2001) and Carney (1979) respectively have parallaxes larger than 10 *mas* and galactic latitude absolutely higher than 30°. Such intermediate –or high– latitude stars which are at distances less than 100 pc can be adopted as free of interstellar extinction hence their UBV–data need no reduction. However there are few stars which do not satisfy both of these conditions though they can be adopted as un–reddened stars. BD +36 2165 ($\pi$ = 8.11 *mas*, b = 67°.35) and HD 39587 ($\pi$ = 115.43 *mas*, b = –02°.73) can be given as two examples for such stars, the galactic latitude of the first star is high and the second star is at a distance of only *r* = 8.7 pc relative to the Sun. 77 stars selected from the two catalogues cited above have been used as "*corner stones*" for the metallicity calibration. Then, 11 stars taken from the catalogue of Cayrel de Strobel et al. (1997) were selected such as to be close to the stars called as "*corner stones*" or to obey the curvature determined by these stars, in the $(\delta_{U-B})_{0.6}$ – [Fe/H] plane. Thus the calibration could be extended down to [Fe/H] = –2.75 dex. No reduction for interstellar extinction was necessary for the UBV–data for five stars in this catalogue which have either large parallaxes or galactic latitudes with $|b| \geq 23°$. Whereas the UBV data for six absolutely very low latitude stars have been de–reddened by the following procedure (Bahcall & Soneira 1980).

---

[1] http://obswww.unige.ch/gcpd/cgi-bin/photoSysHtml.cgi?0

$$A_d(b) = A_\infty(b)[1 - e^{(-\sin b/H)d}] \qquad (1)$$

Here *b* and *d* are the galactic latitude and the distance of the star (evaluated by means of its parallax) respectively. H is the scale-height for the interstellar dust which is adopted as 100 pc and $A_\infty(b)$ and $A_d(b)$ are the total absorptions for the model and for the distance to the star respectively. $A_\infty(b)$ can be evaluated by means of the equation

$$A_\infty(b) = 3.1 E_\infty(B-V) \qquad (2)$$

where $E_\infty(B-V)$ is the colour-excess for the model taken from the NED (NASA Extragalactic Database). Then, $E_d(B-V)$, i.e.: the colour–excess for the corresponding star at the distance *d* can be evaluated by equation (2) adopted for distance *d*

$$E_d(B-V) = A(d)/3.1 \qquad (3)$$

and can be used for the colour excess $E_d(U-B)$, evaluation:

$$E_d(U-B) = 0.72 E_d(B-V) + 0.05 E_d^2(B-V) \qquad (4)$$

Finally, the de–reddened colour indices are:

$$(B-V)_o = (B-V) - E_d(B-V) \quad \text{and} \quad (U-B)_o = (U-B) - E_d(U-B) \qquad (5)$$

### 3. The Method

We adopted the procedure of Carney (1979) for the calibration of the normalized ultraviolet excess relative to Hyades cluster, $(\delta_{U-B})_{0.6}$ and the solar scaled metal abundance [Fe/H], with small modifications. Our sample covers a large range of B-V colour – index, i.e.: $0.37 \leq (B-V)_o \leq 1.07$ mag, however 80% of the stars have colour - indices between 0.40 and 0.70 mag. The normalized ultraviolet excess and the metal abundance cover a large interval, i.e.: $-0.09 \leq (\delta_{U-B})_{0.6} \leq +0.38$ mag and $-2.70 \leq $ [Fe/H] $\leq +0.26$ dex respectively. We divided the interval $-0.09 \leq (\delta_{U-B})_{0.6} \leq +0.38$ mag into 17 scans and adopted the centroid of each scan as a locus point to fit the couple $((\delta_{U-B})_{0.6}, [Fe/H])$. Table 2 gives the locus points and the number of stars associated. It is clear from this table that, the number of metal–poor stars are small in number resulting a relatively larger scale both in $(\delta_{U-B})_{0.6}$ and [Fe/H] encompassing enough stars in this end of the calibration.

A third–degree polynomial is adopted for the locus points (Fig.1). Although the constant term in the equation given by Carney, i.e.: [Fe/H] = $0.11 - 2.90\delta - 18.68\delta^2$ was assumed to represent the metallicity of Hyades and was fixed by 0.11 in the evaluation of the coefficients of other terms. We left it as a free parameter in our calculations. The constant term in the third–degree polynomial resulting by the least–square method is $a_o = 0.10$ which is rather close to the metal abundance given by Carney (1979) and 2% larger than the value of Cameron (1985) i.e.: [Fe/H] = 0.08 dex. The full equation of the polynomial is [Fe/H] = $0.10 - 2.76\delta - 24.04\delta^2 + 30.00\delta^3$. Here $\delta$ is replaced for the normalized ultraviolet excess $(\delta_{U-B})_{0.6}$. The curve of this equation is given in Fig.2 together with the data for all stars in three catalogues cited above. Stars used for the metallicity calibration are marked by a different symbol. The deviations of the evaluated metal abundances from the original ones are given in Fig.3. The mean deviation is almost zero i.e.: <[Fe/H]> = 0.002 dex and the corresponding mean error is ±0.01 dex.

## 4. Discussion

88 dwarfs with solar scaled metallicities –2.70 ≤ [Fe/H] ≤ +0.26 dex have been taken from three different sources for a new metallicity calibration. Especially, the catalogue of Cayrel de Stroble et al. (2001) which provides us 57 stars for this purpose supplies detailed information on abundances for stars with determination based on high–resolution spectroscopy. The selection of these and other 20 stars from the catalogue of Carney (1979) and additional to them (totally 77) have been carried out such as to be free of interstellar extinction. These stars supply at least one of the following conditions: (*i*) the parallax is larger than 10 *mas* (distance relative to the Sun less than 100 pc) and the galactic latitude is absolutely higher than 30$^o$, (*ii*) the parallax is rather large if the galactic latitude is absolutely low and vice versa. The remaining 11 stars were selected from Cayrel de Stroble et al. (1997) by means of a different criterion, i.e.: their metallicities and normalized ultraviolet excesses are in good agreement with the data of 77 stars and their position in the $(\delta_{U-B})_{0.6}$ – [Fe/H] plane obey the curvature determined from 77 stars. Five out of 11 stars could be adopted as free of interstellar extinction whereas the UBV–data of the remaining six stars have to be de–reddened. The last sample extends the metallicity calibration down to [Fe/H] = –2.75 dex.

We compared our results with the ones of Carney (1979). Fig.4 shows the deviations in two works which are evaluated only for the metallicities [Fe/H] ≥ –1.75 dex where the calibration of Carney is valid. There is no discrepancy between two distributions. The mean deviations and mean errors are equal (<[Fe/H]> = 0.00 dex, (*m.e*) = ±0.01 dex). An additional comparison is carried out for 89 metal – poor stars, i.e.: -2.50 ≤ [Fe/H] < -1.75 dex, taken from Carney et al. (1994). The catalogue of these authors contains a larger sample of stars with these metallicities, however many of them are peculiar stars, such as variable stars, binary stars etc. We restricted our sample with a smaller number of stars to avoid any probable error. We used the UBV and E(B-V) data for these stars and evaluated the metallicities by means of new calibration and we compared them with the orginal ones of Carney et al. Fig.5a shows the deviation relative to the orginal metallicities. There is a flat and symmetrical distribution relative to the mean deviation, <Δ[Fe/H]> = 0.21 dex. Contrary to the expectation, the mean deviation is not zero. But, this is also the case for the metal rich stars (Fig.5b), where the mean deviation for metallicity interval -1.75 ≤ [Fe/H] < 0.20 dex is <Δ[Fe/H]> = 0.19 dex, indicating a zero point difference between two sets of data. The mean error for the metal – poor stars, ±0.05 dex, is at the level of expectation. Hence the new calibration provides metallicities with accuracy of Carney's (1979) calibration but it has the advantage of covering the extreme metal–poor stars.

**Acknowledgement:** We thank Prof. Esat Hamzaoğlu for editing the manuscript, and the anonymous referee for his/her comments.

**Table 1.** Dwarfs used for metallicity calibration. The columns give: (1) BD, HD or G- (Giclas) number, (2) Hipparcos number, (3) spectral type, (4) $T_{eff}$, (5) log g, (6), (7), and (8) the UBV data, (9) $\delta_{0.6}$, the standardized ultraviolet excess, (10) [Fe/H], (11) parallax, (12) galactic latitude, and (13) remarks. The figures (1), (2) or (3) in the last column refer to as Cayrel de Stroble et al. (2001), Cayrel de Stroble et al. (1997), and Carney (1979), respectively. The words "corrected" or "uncorrected" denote that UBV data are de-reddened or not (see text). $T_{eff}$ and log g parameters not given in the table of Carney, are from the authors cited in the "remarks" coloumn.

| (1) | (2) | (3) | (4) | (5) | (6) | (7) | (8) | (9) | (10) | (11) | (12) | (13) |
|---|---|---|---|---|---|---|---|---|---|---|---|---|
| No | Hip No | Spec. | $T_{eff}$ | log g | V | B-V | U-B | $\delta_{0.6}$ | [Fe/H] | $\pi$ | *b* | remarks |
| BD +02 0375 | 86443 | A5 | 5793 | 4.00 | 9.820 | 0.420 | -0.260 | 0.36 | -2.50 | 8.35 | 17.03 | (2), corrected |
| BD +09 0352 | 12529 | F2 | 5860 | 4.50 | 10.180 | 0.440 | -0.250 | 0.30 | -2.20 | 5.22 | -44.51 | (1) |
| BD +29 0366 | 10140 | F8V | 5760 | 4.56 | 8.760 | 0.590 | -0.100 | 0.18 | -0.99 | 17.66 | -30.01 | (2), uncorrected |
| BD +36 2165 | 54772 | G0 | 6349 | 4.79 | 9.770 | 0.430 | -0.190 | 0.22 | -1.15 | 8.11 | 67.35 | (1) |
| BD +38 4955 | 114661 | F6 | 5125 | 4.50 | 11.015 | 0.665 | -0.155 | 0.38 | -2.69 | 14.09 | -19.66 | (1) |
| BD +41 3931 | 103269 | G5 | 5560 | 4.77 | 10.170 | 0.590 | -0.130 | 0.25 | -1.60 | 14.24 | -1.82 | (2), corrected |
| BD +42 2667 | 78640 | F5 | 5929 | 4.00 | 9.870 | 0.460 | -0.200 | 0.23 | -1.67 | 8.03 | 48.41 | (3), Rebolo (1988) |
| BD +66 0268 | 16404 | G0 | 5250 | 4.98 | 9.820 | 0.640 | -0.110 | 0.29 | -2.11 | 17.58 | 8.59 | (2), corrected |

Table 1 (Cont.)

| (1) | (2) | (3) | (4) | (5) | (6) | (7) | (8) | (9) | (10) | (11) | (12) | (13) |
|---|---|---|---|---|---|---|---|---|---|---|---|---|
| No | Hip No | Spec. | $T_{eff}$ | log g | V | B-V | U-B | $\delta_{0.6}$ | [Fe/H] | $\pi$ | *b* | remarks |
| BD -06 0855 | 19814 | G:... | 5419 | 4.50 | 10.600 | 0.690 | 0.115 | 0.13 | -0.70 | 24.27 | -37.12 | (1) |
| CD -45 03283 | 36818 | G8V-VI | 5672 | 4.57 | 10.470 | 0.610 | -0.020 | 0.16 | -0.83 | 15.32 | -11.98 | (1) |
| G 88 - 10 | 34630 | A: | 5900 | 4.00 | 11.710 | 0.390 | -0.280 | 0.35 | -2.70 | 4.00 | 14.77 | (2), corrected |
| HD 001581 | 1599 | F9V | 6009 | 4.52 | 4.220 | 0.580 | 0.010 | 0.09 | -0.26 | 116.38 | -51.92 | (1) |
| HD 003765 | 3206 | K2V | 5091 | 4.64 | 7.360 | 0.940 | 0.700 | -0.01 | -0.06 | 57.90 | -22.64 | (2), uncorrected |
| HD 006582 | 5336 | G5Vb | 5305 | 4.61 | 5.170 | 0.700 | -0.100 | 0.16 | -0.71 | 132.42 | -7.87 | (2), uncorrected |
| HD 008673 | 6702 | F7V | 6380 | 4.50 | 6.330 | 0.460 | -0.010 | 0.02 | 0.16 | 26.14 | -27.75 | (1) |
| HD 010700 | 8102 | G8 V | 5500 | 4.32 | 3.500 | 0.720 | 0.210 | 0.08 | -0.36 | 274.18 | -73.44 | (3), Mallik (1998) |
| HD 013555 | 10306 | F5 V | 6358 | 4.07 | 5.290 | 0.420 | -0.070 | 0.09 | -0.40 | 33.19 | -37.81 | (3), Edvardsson (1993) |
| HD 020766 | 15330 | G2.5V | 5860 | 4.50 | 5.520 | 0.630 | 0.080 | 0.08 | -0.20 | 82.51 | -47.21 | (1) |
| HD 022879 | 17147 | F9V | 5926 | 4.57 | 6.700 | 0.540 | -0.080 | 0.15 | -0.76 | 41.07 | -43.12 | (1) |
| HD 028946 | 21272 | K0 | 5288 | 4.55 | 7.930 | 0.770 | 0.360 | 0.02 | -0.03 | 37.33 | -27.24 | (1) |
| HD 030495 | 22263 | G3V | 6000 | 4.50 | 5.470 | 0.600 | 0.140 | -0.01 | 0.10 | 75.10 | -34.81 | (1) |

Table 1 (Cont.)

| (1) | (2) | (3) | (4) | (5) | (6) | (7) | (8) | (9) | (10) | (11) | (12) | (13) |
|---|---|---|---|---|---|---|---|---|---|---|---|---|
| No | Hip No | Spec. | $T_{eff}$ | log g | V | B-V | U-B | $\delta_{0.6}$ | [Fe/H] | $\pi$ | *b* | remarks |
| HD 030649 | 22596 | G1 V-VI | 5727 | 4.31 | 6.970 | 0.590 | 0.020 | 0.11 | -0.32 | 33.44 | 1.02 | (3), Thevenin (1999) |
| HD 039587 | 27913 | G0V | 5929 | 4.50 | 4.410 | 0.590 | 0.080 | 0.03 | -0.05 | 115.43 | -02.73 | (1) |
| HD 052298 | 33495 | F5/F6V | 6072 | 4.60 | 6.940 | 0.460 | -0.110 | 0.14 | -0.84 | 27.38 | -20.34 | (1) |
| HD 056513 | 35377 | G2V | 5659 | 4.50 | 8.030 | 0.630 | 0.050 | 0.11 | -0.38 | 28.19 | 17.57 | (1) |
| HD 063077 | 37853 | G0V | 5820 | 4.42 | 5.360 | 0.570 | -0.070 | 0.17 | -0.80 | 65.79 | -4.81 | (3), Castro (1999) |
| HD 064090 | 38541 | sdG2 | 5370 | 4.00 | 8.260 | 0.610 | -0.120 | 0.26 | -1.73 | 35.29 | 25.93 | (3), Mishenina (2000) |
| HD 064090 | 38541 | sdG2 | 5340 | 4.75 | 8.320 | 0.620 | -0.140 | 0.28 | -1.86 | 35.29 | 25.93 | (2), uncorrected |
| HD 064606 | 38625 | G8V | 5206 | 4.57 | 7.440 | 0.730 | 0.160 | 0.17 | -0.93 | 52.01 | 13.34 | (1) |
| HD 065907 | 38908 | G0V | 6072 | 4.50 | 5.610 | 0.570 | -0.010 | 0.10 | -0.36 | 61.76 | -15.68 | (1) |
| HD 072905 | 42438 | G1.5Vb | 6030 | 4.66 | 5.640 | 0.620 | 0.070 | 0.08 | -0.27 | 70.07 | 35.70 | (3), Gray (2001) |
| HD 074000 | 42592 | sdF6 | 6072 | 4.20 | 9.620 | 0.430 | -0.230 | 0.28 | -2.05 | 7.26 | 15.31 | (3), Hartmann (1988) |
| HD 074000 | 42592 | sdF6 | 6072 | 4.20 | 9.580 | 0.390 | -0.270 | 0.31 | -2.06 | 7.26 | 15.31 | (2), corrected |
| HD 076151 | 43726 | G2V | 5727 | 4.50 | 6.000 | 0.670 | 0.220 | 0.00 | 0.07 | 58.50 | 24.16 | (1) |

Table 1 (Cont.)

| (1) | (2) | (3) | (4) | (5) | (6) | (7) | (8) | (9) | (10) | (11) | (12) | (13) |
|---|---|---|---|---|---|---|---|---|---|---|---|---|
| No | Hip No | Spec. | $T_{eff}$ | log g | V | B-V | U-B | $\delta_{0.6}$ | [Fe/H] | $\pi$ | $b$ | remarks |
| HD 084937 | 48152 | sdF5 | 6222 | 4.00 | 8.320 | 0.370 | -0.200 | 0.27 | -2.19 | 12.44 | 45.47 | (3), Peterson (1981) |
| HD 089125 | 50384 | F8Vbw | 6143 | 4.54 | 5.820 | 0.500 | -0.050 | 0.09 | -0.38 | 44.01 | 55.00 | (1) |
| HD 090508 | 51248 | F9V | 5802 | 4.35 | 6.420 | 0.600 | 0.050 | 0.08 | -0.23 | 42.45 | 54.92 | (3), Fuhrmann (2000) |
| HD 094028 | 53070 | F4V | 6060 | 4.54 | 8.240 | 0.470 | -0.170 | 0.21 | -1.38 | 19.23 | 61.77 | (1) |
| HD 101501 | 56997 | G8V | 5538 | 4.69 | 5.310 | 0.720 | 0.280 | 0.01 | 0.03 | 104.81 | 73.32 | (1) |
| HD 106516 | 59750 | F5V | 6222 | 4.50 | 6.100 | 0.480 | -0.110 | 0.15 | -0.82 | 44.34 | 51.54 | (1) |
| HD 108177 | 60632 | sdF5 | 6200 | 4.40 | 9.670 | 0.430 | -0.220 | 0.26 | -1.70 | 10.95 | 63.42 | (3), Fulbright (2000) |
| HD 110897 | 62207 | G0V | 5860 | 4.41 | 5.950 | 0.550 | -0.030 | 0.11 | -0.31 | 57.57 | 77.78 | (3), Thevenin (1999) |
| HD 113083 | 63559 | F9V | 5750 | 4.50 | 8.050 | 0.550 | -0.110 | 0.19 | -0.93 | 18.51 | 35.44 | (1) |
| HD 114710 | 64394 | F9.5V | 6146 | 4.52 | 4.260 | 0.580 | 0.080 | 0.02 | 0.06 | 109.23 | 85.41 | (1) |
| HD 114762 | 64426 | F9V | 5928 | 4.18 | 7.300 | 0.520 | -0.080 | 0.14 | -0.64 | 24.65 | 79.25 | (3), Clementini (1999) |
| HD 115617 | 64924 | G5V | 5600 | 4.50 | 4.753 | 0.697 | 0.261 | 0.00 | -0.02 | 117.30 | 44.09 | (1) |
| HD 125072 | 69972 | K3V | 4941 | 4.50 | 6.640 | 1.040 | 0.950 | -0.06 | 0.26 | 84.50 | 01.61 | (1) |

Table 1 (Cont.)

| (1) | (2) | (3) | (4) | (5) | (6) | (7) | (8) | (9) | (10) | (11) | (12) | (13) |
|---|---|---|---|---|---|---|---|---|---|---|---|---|
| No | Hip No | Spec. | $T_{eff}$ | log g | V | B-V | U-B | $\delta_{0.6}$ | [Fe/H] | $\pi$ | $b$ | remarks |
| HD 126681 | 70681 | G3V | 5500 | 4.63 | 9.300 | 0.600 | -0.100 | 0.23 | -1.45 | 19.16 | 38.86 | (1) |
| HD 128620 | 71683 | G2V | 5793 | 4.50 | 0.020 | 0.657 | 0.230 | -0.03 | 0.20 | 742.24 | -00.68 | (1) |
| HD 128621 | 71681 | K1V | 5305 | 4.50 | 1.390 | 0.871 | 0.590 | -0.01 | 0.14 | 742.22 | -00.68 | (1) |
| HD 131653 | 72998 | G5 | 5356 | 4.65 | 9.520 | 0.720 | 0.160 | 0.15 | -0.63 | 20.29 | 42.99 | (1) |
| HD 132142 | 73005 | K1V | 5091 | 4.50 | 7.760 | 0.790 | 0.330 | 0.10 | -0.55 | 41.83 | 55.04 | (1) |
| HD 134439 | 74235 | K0/K1V | 5106 | 4.74 | 9.090 | 0.760 | 0.180 | 0.22 | -1.30 | 34.14 | 34.99 | (1) |
| HD 136352 | 75181 | G4V | 5478 | 4.18 | 5.660 | 0.640 | 0.060 | 0.12 | -0.49 | 68.70 | 7.38 | (3), Francois (1986) |
| HD 148816 | 80837 | F8V | 5923 | 4.16 | 7.280 | 0.530 | -0.070 | 0.14 | -0.63 | 24.34 | 33.05 | (3), Clementini (1999) |
| HD 151044 | 81800 | F8V | 6146 | 4.50 | 6.470 | 0.540 | 0.020 | 0.04 | -0.01 | 34.00 | 40.89 | (1) |
| HD 152792 | 82636 | G0V | 5647 | 4.12 | 6.810 | 0.650 | 0.080 | 0.11 | -0.38 | 21.13 | 39.13 | (3), Gorgas (1999) |
| HD 157089 | 84905 | F9V | 5885 | 4.00 | 6.970 | 0.560 | -0.010 | 0.10 | -0.54 | 25.88 | 20.68 | (3), Friel (1992) |
| HD 165908 | 88745 | F7V | 6001 | 4.21 | 5.050 | 0.520 | -0.080 | 0.13 | -0.46 | 63.88 | 22.30 | (3), Gratton (1996) |
| HD 166913 | 89554 | F6:Vw | 6175 | 4.61 | 8.200 | 0.460 | -0.200 | 0.24 | -1.44 | 16.09 | -18.88 | (1) |

Table 1 (Cont.)

| (1) | (2) | (3) | (4) | (5) | (6) | (7) | (8) | (9) | (10) | (11) | (12) | (13) |
|---|---|---|---|---|---|---|---|---|---|---|---|---|
| No | Hip No | Spec. | $T_{eff}$ | log g | V | B-V | U-B | $\delta_{0.6}$ | [Fe/H] | $\pi$ | *b* | remarks |
| HD 181743 | 95333 | F3/F5w | 5929 | 4.25 | 9.660 | 0.460 | -0.250 | 0.31 | -2.04 | 11.31 | -24.27 | (2), uncorrected |
| HD 184960 | 96258 | F7V | 6222 | 4.50 | 5.740 | 0.480 | 0.000 | 0.02 | -0.13 | 39.08 | 14.59 | (1) |
| HD 186185 | 97063 | F5V | 6462 | 4.50 | 5.490 | 0.465 | 0.025 | -0.02 | 0.02 | 27.26 | -18.36 | (1) |
| HD 186427 | 96901 | G3V | 5860 | 4.50 | 6.220 | 0.660 | 0.200 | 0.00 | 0.08 | 46.70 | 13.20 | (1) |
| HD 188510 | 98020 | G5Vw | 5628 | 5.16 | 8.830 | 0.600 | -0.090 | 0.22 | -1.37 | 25.32 | -08.92 | (1) |
| HD 191195 | 99026 | F5V | 6632 | 4.50 | 5.820 | 0.415 | -0.030 | 0.04 | 0.02 | 27.43 | 11.18 | (1) |
| HD 191408 | 99461 | K3V | 4893 | 4.50 | 5.310 | 0.850 | 0.430 | 0.14 | -0.58 | 165.27 | -30.92 | (1) |
| HD 192985 | 99889 | F5V: | 6545 | 4.50 | 5.870 | 0.400 | -0.040 | 0.05 | -0.05 | 28.97 | 05.80 | (1) |
| HD 193901 | 100568 | F7V | 5810 | 4.83 | 8.650 | 0.540 | -0.130 | 0.20 | -1.22 | 22.88 | -29.38 | (1) |
| HD 194598 | 100792 | F7V-VI | 5950 | 4.64 | 8.800 | 0.470 | -0.140 | 0.17 | -0.99 | 17.94 | -16.13 | (2), corrected |
| HD 197039 | 102029 | F5 | 6545 | 4.50 | 6.740 | 0.445 | 0.025 | -0.02 | 0.15 | 14.36 | -15.66 | (1) |
| HD 197373 | 102011 | F6IV | 6462 | 4.50 | 5.990 | 0.420 | -0.040 | 0.05 | -0.03 | 30.12 | 11.33 | (1) |
| HD 197692 | 102485 | F5V | 6632 | 4.50 | 4.138 | 0.427 | 0.010 | -0.01 | -0.11 | 68.16 | -35.50 | (1) |

Table 1 (Cont.)

| (1) | (2) | (3) | (4) | (5) | (6) | (7) | (8) | (9) | (10) | (11) | (12) | (13) |
|---|---|---|---|---|---|---|---|---|---|---|---|---|
| No | Hip No | Spec. | $T_{eff}$ | log g | V | B-V | U-B | $\delta_{0.6}$ | [Fe/H] | $\pi$ | $b$ | remarks |
| HD 197963 | 102531 | F7V | 6300 | 4.50 | 5.140 | 0.490 | 0.080 | -0.06 | 0.12 | 31.69 | -16.58 | (1) |
| HD 199289 | 103498 | F5V | 5936 | 4.71 | 8.290 | 0.520 | -0.130 | 0.19 | -0.99 | 18.94 | -40.65 | (1) |
| HD 201891 | 104659 | F8V-VI | 5867 | 4.46 | 7.370 | 0.510 | -0.160 | 0.21 | -1.42 | 28.26 | -20.43 | (3), Edvardsson (1993) |
| HD 202628 | 105184 | G2V | 5771 | 4.52 | 6.740 | 0.630 | 0.130 | 0.03 | -0.14 | 42.04 | -44.45 | (1) |
| HD 204121 | 105864 | F5V | 6545 | 4.50 | 6.120 | 0.450 | -0.010 | 0.02 | 0.08 | 20.89 | -33.25 | (1) |
| HD 210752 | 109646 | G0 | 5958 | 4.59 | 7.400 | 0.520 | -0.080 | 0.14 | -0.59 | 26.57 | -47.05 | (1) |
| HD 212698 | 110778 | G3V | 5915 | 4.50 | 5.540 | 0.610 | 0.060 | 0.08 | -0.13 | 49.80 | -54.98 | (1) |
| HD 212754 | 110785 | F7V | 6146 | 4.50 | 5.760 | 0.515 | 0.030 | 0.01 | -0.04 | 25.34 | -42.93 | (1) |
| HD 213042 | 110996 | K5V | 4760 | 4.58 | 7.670 | 1.070 | 1.000 | -0.09 | 0.25 | 64.74 | -58.77 | (1) |
| HD 217014 | 113357 | G2.5IVa | 5669 | 4.06 | 5.500 | 0.670 | 0.200 | 0.01 | 0.12 | 65.10 | -34.73 | (3), Gratton (1996) |
| HD 217877 | 113896 | F8V | 6000 | 4.50 | 6.680 | 0.580 | 0.060 | 0.04 | -0.10 | 32.50 | -56.02 | (1) |
| HD 218235 | 114081 | F6Vs | 6462 | 4.50 | 6.160 | 0.445 | 0.020 | -0.01 | 0.25 | 23.16 | -37.72 | (1) |
| HD 218261 | 114096 | F7V | 6146 | 4.50 | 6.450 | 0.540 | 0.020 | 0.04 | 0.09 | 35.32 | -36.54 | (1) |

Table 1 (Cont.)

| (1) | (2) | (3) | (4) | (5) | (6) | (7) | (8) | (9) | (10) | (11) | (12) | (13) |
|---|---|---|---|---|---|---|---|---|---|---|---|---|
| No | Hip No | Spec. | $T_{eff}$ | log g | V | B-V | U-B | $\delta_{0.6}$ | [Fe/H] | $\pi$ | *b* | remarks |
| HD 218470 | 114210 | F5V | 6545 | 4.50 | 5.600 | 0.405 | -0.035 | 0.04 | -0.17 | 29.33 | -10.16 | (1) |
| HD 222451 | 116824 | F1V | 6632 | 4.50 | 6.250 | 0.400 | -0.010 | 0.01 | 0.09 | 22.63 | -24.02 | (1) |

**Table 2.** Locus points and the number of stars associated with them (last column). The other columns give the current number, $\delta_{0.6}$, [Fe/H], mean errors for the $\delta_{0.6}$ and [Fe/H], respectively.

| No | $\delta_{0.6}$ | [Fe/H] | $\Delta\delta_{0.6}$ | $\Delta$[Fe/H] | N |
|----|------|-------|------|------|---|
| 01 | −0.07 | +0.21 | 0.01 | 0.04 | 3 |
| 02 | −0.02 | +0.09 | 0.00 | 0.04 | 8 |
| 03 | +0.01 | +0.05 | 0.00 | 0.02 | 7 |
| 04 | +0.02 | +0.01 | 0.00 | 0.04 | 7 |
| 05 | +0.04 | −0.04 | 0.00 | 0.03 | 7 |
| 06 | +0.08 | −0.28 | 0.00 | 0.03 | 8 |
| 07 | +0.11 | −0.41 | 0.00 | 0.03 | 7 |
| 08 | +0.14 | −0.62 | 0.00 | 0.04 | 8 |
| 09 | +0.15 | −0.75 | 0.00 | 0.03 | 5 |
| 10 | +0.17 | −0.93 | 0.00 | 0.04 | 4 |
| 11 | +0.19 | −1.05 | 0.00 | 0.07 | 3 |
| 12 | +0.22 | −1.32 | 0.00 | 0.04 | 5 |
| 13 | +0.23 | −1.52 | 0.00 | 0.06 | 3 |
| 14 | +0.26 | −1.68 | 0.00 | 0.03 | 3 |
| 15 | +0.28 | −2.05 | 0.00 | 0.06 | 4 |
| 16 | +0.31 | −2.10 | 0.00 | 0.04 | 3 |
| 17 | +0.36 | −2.60 | 0.01 | 0.05 | 3 |

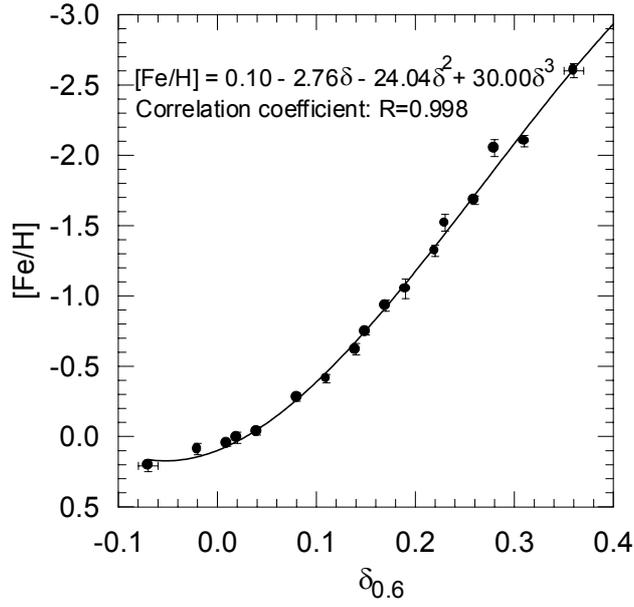

**Figure 1.** The third–degree polynomial curve throught 17 locus–points and the correlation coefficient. The bars show the mean errors.

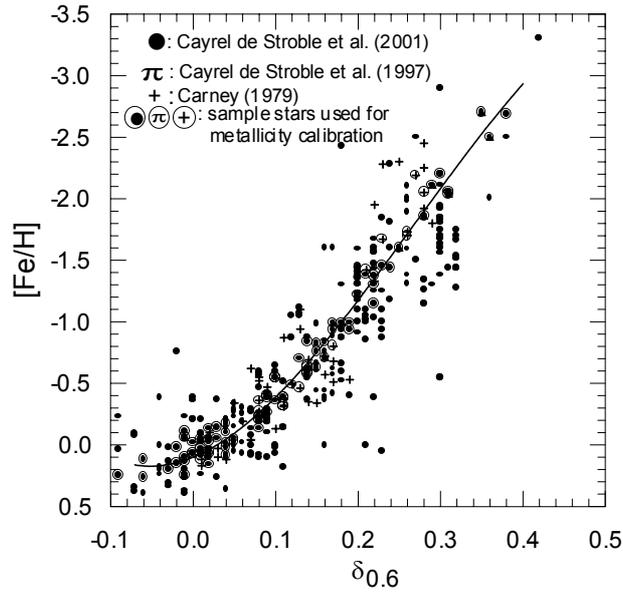

**Figure 2.** The third – degree polynomial curve evaluated by means of 17 locus – points of 88 dwarfs, selected from three catalogues defined by the criteria explained in the text, and the position of stars which do not satisfy the mentioned criteria (see the text). The symbols give: (•) dwarfs with log g ≥ 4.5 from Cayrel de Stroble et al. (2001), (π) dwarfs with log g ≥ 4.0 from Cayrel de Stroble et al. (1997), and (+) dwarfs from Carney (1979). A circled star belongs to the sample of 88 stars.

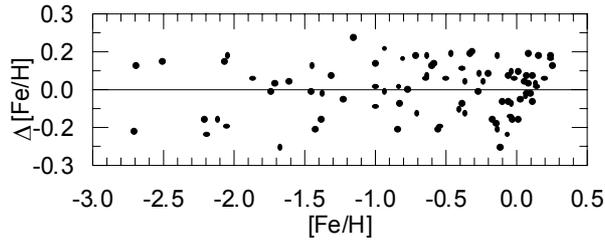

**Figure 3.** Deviation of evaluated metallicities from original ones versus orginal metallicity. The mean deviation and the mean error for this distribution are <[Fe/H]> = 0.00 and (m.e.) = ± 0.01 dex, respectively.

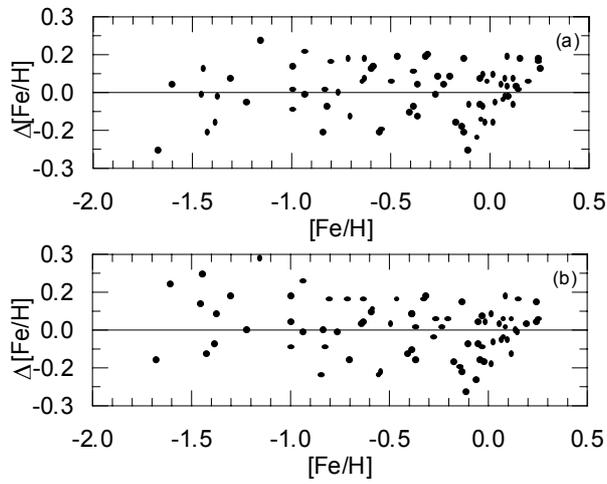

**Figure 4.** Comparison of the deviations in our work (a) and in the work of Carney (b) for [Fe/H] ≥ -1.75 dex where Carney's calibration is valid. There is no discrepancy between two distributions. Also, the mean deviations and the mean errors in these works are equal, respectively.

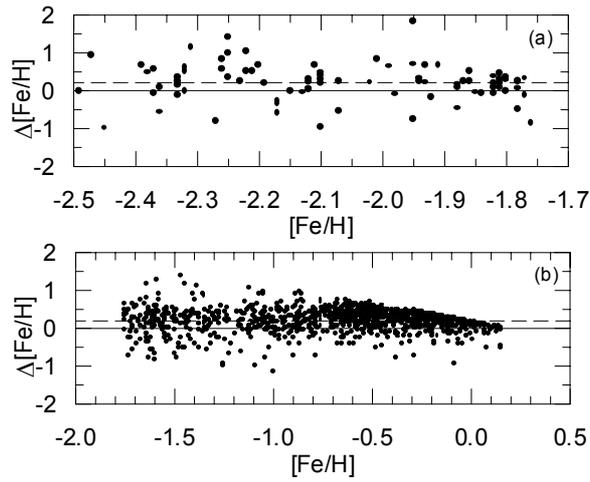

**Figure 5.** Deviations of the evaluated metallicities relative to the original ones, taken from Carney et al. (1994). (a) for metallicities -2.50 ≤ [Fe/H] < -1.75 dex, and (b) for -1.75 ≤ [Fe/H] < 0.20 dex. The mean deviations, different than zero (dashed lines), are due to different zero points in two systems.